\documentclass[aps,prl,showpacs, 
twocolumn] 
{revtex4}
\usepackage{graphicx} 
\usepackage{hyperref} 
\usepackage{amssymb, amsmath}
\usepackage{epsf}

\def\la{\; \raise0.3ex\hbox{$<$\kern-0.75em\raise-1.1ex\hbox{$\sim$}}\;}
\def\ga{\;  \raise0.3ex\hbox{$>$\kern-0.75em\raise-1.1ex\hbox{$\sim$}}\;}

\begin{document}

\title{New class of g-modes 
and unexpected convection in neutron stars}
%
%
\author{Mikhail E. Gusakov$^{1}$ and Elena M. Kantor$^{1,2}$}
\affiliation{
$^1$Ioffe Physical Technical Institute,
Polytekhnicheskaya 26, 194021 St.-Petersburg, Russia
\\
$^2$St.-Petersburg State Polytechnical University,
Polytekhnicheskaya 29, 195251 St.-Petersburg, Russia
}

\begin{abstract}
We suggest a specific new class  
of low-frequency g-modes
in superfluid neutron stars.
We determine the Brunt-V$\ddot{\rm a}$is$\ddot{\rm a}$l$\ddot{\rm a}$ frequency
for these modes 
and demonstrate that they 
can be unstable with respect to convection.
The criterion for the instability onset 
(analogue of the well known Schwarzschild criterion)
is derived. 
It is very sensitive to 
equation of state and a model of nucleon superfluidity.
In particular, convection may occur
for both positive and negative temperature gradients.
Our results have interesting implications 
for neutron star cooling and seismology.
\end{abstract}

\maketitle

{\it Introduction}. ---
This Letter is devoted to {\it gravity} oscillation modes (g-modes) 
and the related phenomenon of convection in neutron stars (NSs).
The restoring force for g-modes is buoyancy 
that originates from the dependence of the pressure 
on at least two quantities 
(e.g., density and temperature or density and chemical composition).
g-modes and convection are actively studied 
in laboratory experiments (e.g., \cite{abfghlsv12, zks08, whw80}) 
and are widespread in nature.
For instance, 
g-modes are observed in Earth atmosphere and ocean, 
in white dwarfs \cite{wk08}, 
in slowly pulsating B-stars \cite{cat07},
and in other objects \cite{cunhaetal07},
while convection is typical for most of the stars
(including the Sun).
In application to NSs, 
g-modes 
were studied, e.g., 
in Refs.\ \cite{mhs83,finn87,dhh88,rg92,lai94,lai99}.
In all these works the NS matter was assumed 
to be nonsuperfluid (normal).
However, according to microscopic calculations \cite{ls01,yls99},
baryons in the internal layers of NSs become
superfluid (SFL) at temperatures $T \la 10^8 \div 10^{10}$~K
which has a drastic impact on stellar dynamics and evolution 
\cite{yls99, ch08}.
Recent real-time observations \cite{hh10} 
of the cooling NS in Cas A supernova remnant
indicate 
that this NS has an SFL core \cite{shternin11,page11}.
A number of attempts \cite{Lee95,ac01,pr02} 
have been made to theoretically predict g-modes
in cold SFL NSs,
but they have failed.
This led to a general belief 
that g-modes do not exist in SFL interiors of NSs
or, more precisely, 
their frequencies $\omega$ are all degenerate at zero.
In this Letter we show that proper account 
of finite temperature effects 
extracts g-modes from the zero frequency domain.
Moreover, 
these modes can be {\it unstable} 
with respect to convection.
Possible applications of these results
are outlined.
Below the Planck constant,
the speed of light,
and the Boltzmann constant
equal unity, 
$\hbar=c=k_{\rm B}=1$.

{\it Convection in 
NSs and SFL g-modes.} ---
%
For simplicity, 
consider $npe$ NS cores, 
composed of 
neutrons ($n$), protons ($p$), and electrons ($e$).
To start with, 
assume that all particles are nonsuperfluid.
Any thermodynamic quantity in $npe$-matter 
(e.g., the heat function  $w=\varepsilon+P$, 
where $\varepsilon$ is the energy density 
and $P$ is the pressure)
can be presented as a function of 3 variables, 
say, $P$, $x_e\equiv n_e/n_b$, 
and $x_S\equiv S/n_b$. 
Here $n_i$ 
is 
the number density 
for particles $i=n$, $p$, and $e$;
$n_b$ is the baryon number density; 
$S$ is the entropy density.
What is the local criterion for the absence of 
convection in $npe$-matter?
It is easy to derive it
in the same manner as it was done in Refs.\ \cite{ll87,thorne66} 
(see also \cite{lai94}).
Assume that a spherically symmetric star 
is in hydrostatic equilibrium 
(but not necessarily in thermal or beta-equilibrium), 
that is
%
$\nabla P=-w \, \nabla \phi$,
%
where 
$\phi(r)$ is the gravitational potential and
$r$ is the radial coordinate.
Here and below 
$\nabla \equiv d/dr$
because all quantities of interest depend on $r$ only.
Consider two close points $1$ and $2$
with $r=r_1$ and $r_2$. 
Let $A_1$ and $A_2$ be the values 
of some thermodynamic quantity $A$
at points $1$ and $2$, respectively,
and
$\Delta A \equiv A_2-A_1$.
Let us displace adiabatically a small fluid element 
upward from point $1$ to point $2$.
At point $2$ the pressure of the fluid element
adjusts itself to the surrounding pressure $P_2=P_1+\Delta P$, 
while $x_e$ and $x_S$ remain 
unchanged
and equal to $x_{e1}$ and $x_{S1}$ 
(we assume that beta-processes are slow).
The matter is stable against convection if
the inertial mass density 
of the lifted element is larger 
than the equilibrium density at point $2$.
For relativistic matter the role of the inertial mass density
is played by $w$ \cite{mt66}. 
Thus, the stability requires that
$w(P_2, \, x_{e2}, \, x_{S2}) < w(P_2, \, x_{e1}, \, x_{S1})$.
Expanding $w$ in Tailor series near point $1$, 
we obtain
\begin{equation}
\partial_{x_e} w(P,\, x_e,\, x_S) \,\, \nabla x_e +
\partial_{x_s} w(P,\, x_e,\, x_S) \,\, \nabla{x_S} <0,
\label{conv1}
\end{equation}
where $\partial_A \equiv \partial/\partial A$.
When $w$ is a function of $P$ and $x_S$ only, 
Eq.\ (\ref{conv1}) immediately 
reproduces the textbook criterion 
for the absence of convection (see, e.g., \cite{ll87}).
In a strongly degenerate matter the second term in Eq.\ (\ref{conv1})
is much smaller than the first one and can be neglected.
Similarly,
to calculate the first term in Eq.\ (\ref{conv1}) 
it is sufficient to 
set $T=0$ and $x_S=0$.
Then, Eq.\ (\ref{conv1}) reduces to
%
$\partial_{x_e} w(P,\, x_e) \,\, \nabla{x_e} <0$.
%
This Ledoux-type criterion is always satisfied 
in beta-equilibrated NSs,
i.e., they are stable against convection.
Oscillations of such a matter near equilibrium correspond to
temperature-independent {\it composition} g-modes, 
first studied in Ref.\ \cite{rg92}.

Assume now
that neutrons (and possibly protons) are SFL.
What will be the analogue of criterion (\ref{conv1}) in that case?
Nucleon SFL 
leads to the appearance of
two independent velocity fields: 
SFL neutron velocity ${\pmb V}_{{\rm s}n}$
and velocity of normal liquid component ${\pmb V}_{\rm q}$,
composed of neutron Bogoliubov excitations, 
protons
and electrons \cite{footnote1}.
The presence of extra velocity field ${\pmb V}_{{\rm s}n}$ 
results in additional 
(besides equation $\nabla P = -w \nabla \phi$)
condition of hydrostatic equilibrium 
in SFL matter \cite{ga06, gkcg12}:
%
$\nabla \left(\mu_n \, {\rm e}^\phi \right)=0$,
%
where $\mu_n$ is 
the relativistic neutron chemical potential. 
As a result, when we displace the fluid element, 
`attached' to the {\it normal} liquid component,
from point $1$ to point $2$,
both $P$ and $\mu_n$ 
adjust themselves 
to their equilibrium values $P_2$ and $\mu_{n2}$ at point $2$.
The pressure 
adjusts by contraction/expansion of the fluid element,
while $\mu_n$ adjusts by the variation in
the number of `SFL neutrons' in this element.
Note that, since SFL neutrons can freely escape from 
the fluid element attached
to the normal particles, 
the total number of neutrons in the element is not conserved,
and neither are the quantities $x_e=n_e/n_b$ and $x_S=S/n_b$.
In this situation the conserved quantity is $x_{eS}=S/n_e$, 
because both 
the entropy and electrons flow with the same velocity ${\pmb V}_{\rm q}$ 
(e.g., \cite{khalatnikov89,ga06,gkcg12}).
Bearing this in mind, it is convenient to 
consider $w$ as a function of $P$, $\mu_n$, and $x_{eS}$.
Then the condition for stability against convection can be written as
$w(P_2,\,\mu_{n2},\,x_{eS2}) < w(P_2,\,\mu_{n2},\,x_{eS1})$ or
\begin{equation}
\partial_{x_{eS}} w(P,\, \mu_n, \, x_{eS}) \,\, \nabla{x_{eS}} <0.
\label{conv2}
\end{equation}
A similar condition was derived in a different way 
in Ref.\ \cite{parshin69}, 
where internal gravity waves were analyzed 
in a mixture of SFL He-4 and a normal fluid 
(see also \cite{steinberg80, fetter82}).
Note that 
the left-hand side of Eq.\ (\ref{conv2}) 
depends on $T$ and vanishes at $T=0$. 
Then the system is marginally stable, 
since there is no restoring force acting on a displaced fluid element.
Thus, it is not surprising 
that the authors of Refs.\ \cite{Lee95,ac01,pr02},
who assumed $T=0$, 
did not find g-modes in SFL NSs.
In contrast, 
consistent treatment of the temperature effects 
should reveal g-modes. 

To check it
we perform a local analysis of SFL hydrodynamic equations
(see, e.g., \cite{ga06,gkcg12}),
describing oscillations 
of an NS in the weak-field limit ($\phi \ll 1$) 
at $T \neq 0$.
We analyze short-wave perturbations, 
proportional to ${\rm exp}(i \omega t)\,{\rm exp}[i \int^r dr' k(r')]\,{\rm Y}_{lm}$,
where the wave number $k$ of a perturbation weakly depends on $r$
($k\gg |d \, {\rm ln} k/dr|$, WKB approximation), 
and ${\rm Y}_{lm}$ is a spherical harmonic.
Solving oscillation equations in the Cowling approximation 
(in which $\phi$ is not perturbed \cite{cowling41}), 
we find the standard \cite{rg92} short-wave dispersion relation 
for the SFL g-modes:
%
$\omega^2 = {\mathcal N}^2 \, l(l+1)/[l (l+1)+k^2 r^2]$,
%
where
\begin{equation}
{\mathcal N}^2 = 
-\frac{g}{\mu_n n_b} \, \frac{(1+y)}{y} \,
\partial_{x_{eS}} w(P,\, \mu_n, \, x_{eS}) \,\, \nabla{x_{eS}}
\label{N2}
\end{equation}
is the corresponding 
Brunt-V$\ddot{\rm a}$is$\ddot{\rm a}$l$\ddot{\rm a}$ 
frequency squared;
$g= \nabla \phi$; 
$y=n_b \, Y_{pp}/[\mu_n \, (Y_{nn}Y_{pp}-Y_{np}^2)]-1 >0$,
$Y_{ik}$ being the relativistic entrainment matrix
(see, e.g., \cite{gkh09a,gkh09b} 
and comment \cite{footnote2}). 
The stability condition for these g-modes, 
$\mathcal{N}^2>0$,
coincides with Eq.\ (\ref{conv2}).

For numerical evaluation of Eq.\ (\ref{N2}) 
it is convenient 
to introduce a new set 
of independent variables 
$n_b$, $n_e$, and $T$ 
instead of 
$P$, $\mu_n$, and $x_{eS}$.
Then Eq.\ (\ref{N2}) is approximately 
rewritten as
\begin{equation}
\mathcal{N}^2\approx 
\alpha \, C_V F
\left[-\nabla{T}/(g T) + F - 1 \right], 
\label{N3}
\end{equation}
where we neglect small 
terms of the second and higher orders in $T/\epsilon_{\rm F}$ 
($\epsilon_{\rm F}$ is the typical particle Fermi energy).
In Eq.\ (\ref{N3}) 
$\alpha =g^2 \, T \, (1+y)/(y \, \mu_n \, n_b) > 0$
and
$C_V= T \, \partial_T S >0$.
Finally, 
$F=1+\mu_n G_2/(G_1 \, C_V)$, where
$G_1 = \partial_{n_e} P \,\, \partial_{n_b} \mu_n -
\partial_{n_b} P  \,\, \partial_{n_e} \mu_n$
and 
$G_2= \partial_{n_e} P \,\, \partial_{n_b} S 
+ \partial_{n_e} \mu_n \,\, 
\left(S - n_b \, \partial_{n_b} S -n_e \, \partial_{n_e} S \right)$.
%
%
For an NS in thermal equilibrium, 
the red-shifted temperature 
$T^\infty \equiv T \, {\rm e}^{\phi}$ 
is constant throughout the core, 
$ \nabla T^\infty = (\nabla T + g T) \, {\rm e}^\phi=0$.
%
%
In that case ${\mathcal N}^2$ in Eq.\ (\ref{N3}) is positive 
and reduces to
\begin{equation}
\mathcal{N}^2 \approx \alpha \, C_V \, F^2.
\label{N4}
\end{equation}
If NS is not in thermal equilibrium, 
$\mathcal{N}^2$ can be negative 
for certain $\nabla T^\infty$. 
These gradients follow from Eq.\ (\ref{N3}) 
(or Eq.\ \ref{conv2})
and are defined by the inequality
%
$F \nabla T^\infty>g T^\infty F^2$,
%
which is the analogue 
of the ordinary Schwarzschild criterion 
for convection \cite{thorne66}. 
This inequality, 
as well as Eq.\ (\ref{conv2}), 
is valid
not only in the weak-field limit $\phi \ll 1$, 
but also in the full general relativity. 
When it
is satisfied,
convective instability occurs \cite{footnote3}. 
Thus, the critical gradient 
for the instability onset is given by
\begin{equation}
 \nabla T_{\rm crit}^\infty = g T^\infty F.
 \label{Tcrit}
\end{equation}
Clearly,
$F$ determines 
the sign of $\nabla T_{\rm crit}^\infty$. 
If $F<0$ one gets the instability while heating 
the matter from below;
if $F>0$ the instability occurs when it is heated from above.
Note that both signs of $\nabla T^\infty$ 
can be realized in cooling NSs \cite{gyp01}.

\begin{figure}[t]
\setlength{\unitlength}{1mm}
\leavevmode
\includegraphics[width=75mm]{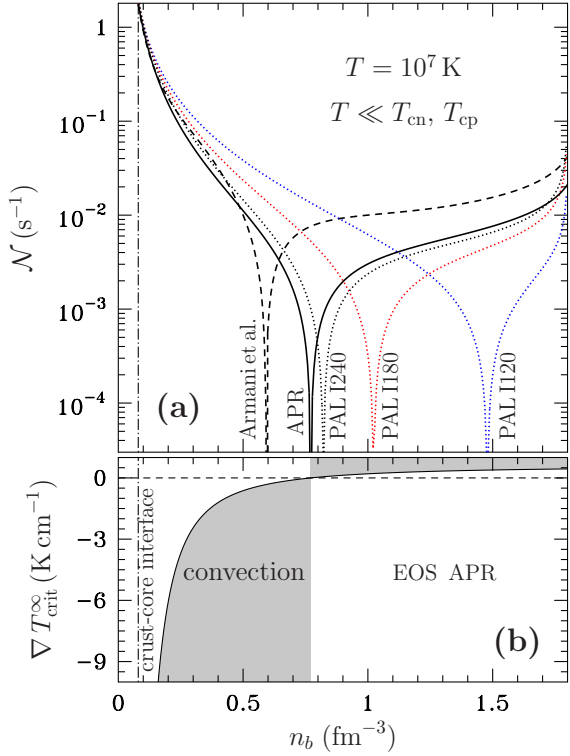}
\caption{
Panel (a): $\mathcal{N}$ 
versus $n_b$ 
for EOSs of Armani et al. \cite{armani11}, APR \cite{apr98}, and PAL \cite{pal88}.
We adopt the model I of PAL family 
with three values of the compression modulus, $120$, $180$ and $240$ MeV.
Panel (b): $\nabla T_{\rm crit}^\infty$ versus $n_b$ for APR EOS. 
The instability region is filled with gray.
Vertical dot-dashed line corresponds to the crust-core interface. 
Both panels are plotted for $T=10^7\,\rm K$. 
Here and in Fig.\ \ref{fig2} 
$g=10^{14}$~cm~s$^{-2}$.
}
\vskip -5mm
\label{fig1}
\end{figure}

{\it Results}. ---
%
First, consider the limit 
$T\ll T_{{\rm c}n},\,T_{{\rm c}p}$
($T_{{\rm c}i}$ is the critical temperature 
for particles $i=n,\,p$), 
in which the nucleon entropy is
exponentially suppressed
and the entropy density $S$ 
is provided by electrons, 
$S=S_e=T (3\pi^2 n_e)^{2/3}/3$.
In this limit 
both $\mathcal{N}$ and $\nabla T_{\rm crit}^{\infty}$ 
are $\propto T$.
Fig.\ \ref{fig1}(a) presents 
the Brunt-V$\ddot{\rm a}$is$\ddot{\rm a}$l$\ddot{\rm a}$ 
frequency $\mathcal{N}$
(see Eq.\ \ref{N4})
as a function of $n_b$ 
for $npe$-matter in thermodynamic equilibrium
for 5 equations of state (EOSs) and $T=10^7$~K. 
One sees that for any EOS 
$\mathcal{N}$ vanishes 
at a certain $n_b$
that corresponds to $F=0$. 
Fig.\ \ref{fig1}(b) shows the critical gradient 
$\nabla T_{\rm crit}^\infty$
(see Eq.\ \ref{Tcrit})
versus $n_b$ for the APR EOS \cite{apr98}.
As expected, 
$\nabla T_{\rm crit}^\infty$ 
changes sign when $F=0$. 
The region of parameters, 
where convection occurs, 
is filled with gray. 

\begin{figure}[t]
\setlength{\unitlength}{1mm}
\leavevmode
\includegraphics[width=72mm]{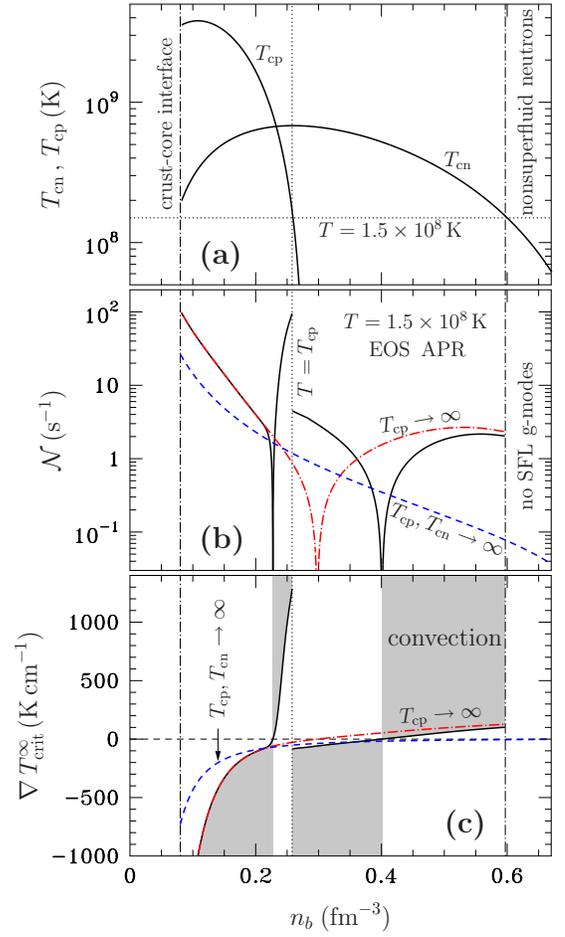}
\caption{
Panel (a): $T_{{\rm c}n}$ and $T_{{\rm c}p}$ versus $n_b$. 
Panels (b) and (c): $\mathcal{N}$ and $\nabla T_{\rm crit}^\infty$ versus $n_b$
for EOS APR and $T=1.5 \times 10^8$~K 
[see the horizontal dotted line in panel (a)].
Solid lines in panels (b) and (c) are obtained for
$T_{{\rm c}n}(n_b)$ and $T_{{\rm c}p}(n_b)$ from panel (a);
dot-dashed lines: $T_{{\rm c}n}(n_b)$ is from panel (a),
$T_{{\rm c}p} \rightarrow \infty$;
dashed lines: $T_{{\rm c}n},\,T_{{\rm c}p}\rightarrow \infty$.
The right vertical dot-dashed line indicates the boundary between the SFL
and normal neutron matter (in the latter SFL g-modes are absent).
The vertical dotted line shows a similar boundary for protons.
Other notations are the same as in Fig.\ \ref{fig1}.
}
\vskip -5mm
\label{fig2}
\end{figure}

When $T$ is not too low 
($ 0.1 T_{{\rm c}i} \la T \la T_{{\rm c}i}$, $i=n$ and/or $p$), 
nucleonic contribution to $S$ is non-negligible.
The results in that case strongly 
differ from those obtained in the limit 
$T \ll T_{{\rm c}n}, \, T_{{\rm c}p}$
and are presented in Fig.\ \ref{fig2}. 
For illustration, we adopt the APR EOS
and take $T=1.5\times 10^8\, \rm K$. 
Some realistic profiles of 
singlet proton $T_{{\rm c}p}(n_b)$
and triplet neutron $T_{{\rm c}n}(n_b)$
critical temperatures,
employed in our numerical calculations, 
are shown in Fig.\ \ref{fig2}(a). 
Figs.\ \ref{fig2}(b,c) demonstrate
$\mathcal{N}(n_b)$  and 
$\nabla T_{\rm crit}^\infty(n_b)$, respectively. 
Solid lines in Figs.\ \ref{fig2}(b,c) 
are obtained for
$T_{{\rm c}n}(n_b)$ and $T_{{\rm c}p}(n_b)$ 
from Fig.\ \ref{fig2}(a).
Dot-dashed lines are plotted 
for $T_{{\rm c}n}(n_b)$ from Fig.\ \ref{fig2}(a), 
but for $T_{{\rm c}p}\rightarrow \infty$.
Finally, dashed lines in Figs.\ \ref{fig2}(b,c) correspond to 
the limit
$T_{{\rm c}n},\,T_{{\rm c}p} \rightarrow \infty$
of Fig.\ \ref{fig1}
(note, however, that Fig.\ \ref{fig1} 
is plotted for different $T$). 
%
%
For simplicity, 
we assume that the entropy density $S_i$ 
of particles $i=n$, $p$, and $e$ depends only on $n_i$ and $T$, 
so that $S=S_n(n_n,T)+S_p(n_p,T)+S_e(n_e,T)$.
Then $F$, which enters Eqs.\ (\ref{N4}) and (\ref{Tcrit}), 
can be rewritten as
\begin{eqnarray}
&&F=1+\frac{\mu_n}{G_1 C_V} 
\left[\left(\partial_{n_e} P -n_n \, \partial_{n_e} \mu_n \right) 
\partial_{n_n} S_n
\right.
\nonumber \\
 && \left.
 -n_e \, \partial_{n_e} \mu_n \,\, 
 \partial_{n_p} S_p - n_e \, \partial_{n_e} \mu_n \,\, 
 \partial_{n_e} S_e + S \, \partial_{n_e} \mu_n \right].\quad \,\,\,
\label{F}
\end{eqnarray}
Clearly, the behaviour of $\mathcal{N}$ 
and $\nabla T_{\rm crit}^\infty$ 
results from 
interplay between 
various derivatives 
of $S_i$.
The key role is played by
the {\it nucleon} derivatives
$\partial_{n_i} S_i$ ($i=n, \, p$).
Because $S_i$ 
strongly depends on $T_{{\rm c}i}$,
which is in turn a very strong function of $n_b$ 
(especially, on the slopes of $T_{{\rm c}i}(n_b)$, 
see Fig.\ \ref{fig2}a),
$\partial_{n_i} S_i$
can be {\it very large} and thus determine the dependence of 
${\mathcal N}$ and $\nabla T_{\rm crit}^\infty$ on $n_b$.
 
To illustrate this point
let us 
compare dot-dashed and dashed curves
in Fig.\ \ref{fig2}(c). 
Dashed curve is plotted assuming $S_n=S_p=0$ 
($T_{{\rm c}n}, \, T_{{\rm c}p} \rightarrow \infty$)
while for dot-dashed curve only $S_p=0$.
The terms in $F$ related to 
$\partial_{n_n} S_n$
are negative for $n_b \la 0.2$~fm$^{-3}$
and positive for $n_b \ga 0.2$~fm$^{-3}$.
This can be easily understood if one bears in mind
that for densities of interest 
($i$) $G_1>0$ and
$\partial_{n_e}P-n_n \partial_{n_e} \mu_n>0$;
($ii$) $S_n$ decreases with increasing $T_{{\rm c}n}$;
and 
($iii$) $T_{{\rm c}n}$ reaches maximum 
at $n_b \sim 0.2$~fm$^{-3}$ 
(Fig.\ \ref{fig2}a).
As a result, 
$\nabla T_{\rm crit}^\infty$ 
given by the dot-dashed curve 
vanishes at lower $n_b$ than for dashed curve (Fig.\ \ref{fig2}c).

Now let us consider the effects related to
the proton entropy density $S_p$
(solid line in Fig.\ \ref{fig2}c). 
When $T \ll T_{{\rm c}p}$ 
the solid and dot-dashed curves coincide, 
because in that case $S_p$ is negligible. 
When $T$ approaches $T_{{\rm c}p}$ with growing $n_b$, 
$\nabla T_{\rm crit}^\infty$ 
rapidly increases because 
$dT_{{\rm c}p}/dn_b$ is large and negative 
(i.e. $\partial_{n_p} S_p>0$)
and $\partial_{n_e} \mu_n <0$ 
[see the corresponding term in Eq.\ (\ref{F})].
At $T=T_{{\rm c}p}$ 
(vertical dotted line in Fig.\ \ref{fig2}), 
$\partial_{n_p} S_p$ and $\partial_T S_p$ 
are discontinuous that results in discontinuities of 
$\mathcal{N}$ 
and $\nabla T_{\rm crit}^\infty$. 
At $n_b \ga 0.26$~fm$^{-3}$ all protons are normal
and $S_p=T\, m_p^\ast \,p_{{\rm F}p}/3$, 
where $m_p^\ast$ and $p_{{\rm F}p}$ 
are the proton effective mass and 
Fermi momentum, respectively.
At such $n_b$
the proton contribution to $F$ 
is negative, i.e. 
the solid curve in Fig.\ \ref{fig2}(c) goes lower 
than the dot-dashed curve.
The most important conclusion drawn 
from the analysis 
of Fig.\ \ref{fig2} 
is that SFL g-modes and
convection in the internal layers of NSs 
are {\it extremely sensitive}
to the EOS and the model of nucleon SFL.
An account for the singlet neutron SFL 
at lower densities $n_b \la 0.08$~fm$^{-3}$
may additionally affect ${\mathcal N}$ 
and $\nabla T_{\rm crit}^\infty$
near the crust-core interface.

{\it Discussion and conclusion}. ---
%
Our results indicate 
that NSs can have convective internal layers.
This could affect the thermal evolution of young NSs 
(such as in Cas A), 
for which $\nabla T^\infty$ is not completely smoothed out 
by the thermal conductivity, 
as well as the thermal relaxation 
of quasi-persistent X-ray transients \cite{syhp07, bc09}.
Note that in this Letter we only consider
SFL g-modes and convection in the NS cores 
(but not in the crust).
If the NS crust is {\it elastic}, 
as it is usually assumed, 
then it is most likely that core SFL g-modes
do not penetrate the crust, 
while the crustal SFL g-modes
are `mixed' with the shear modes \cite{dhh88}
(for which the restoring force is elasticity),
and pushed to frequencies 
$\omega \gg {\mathcal N}$ 
(but see comment \cite{footnote5}).
In that case convection is absent.
However, if the inner crust 
(especially, mantle \cite{ch08}) 
is {\it plastic} \cite{ll12} 
then the existence of crustal SFL g-modes 
and convection cannot be excluded,
that can have even more interesting implications 
for NS cooling.

We have shown that g-modes
can propagate in SFL NS matter. 
But how can they be excited?
Among the potential scenarios is the excitation 
of stable SFL g-modes by unstable ones 
(i.e., by convective motions).
Another possibility was considered in Refs.\ \cite{lai94,hl99} 
in application to composition g-modes of normal NSs.
It consists in resonant excitations of SFL g-modes
by tidal interaction in coalescing binary systems, 
when the frequency of the tidal driving force 
equals one of the g-mode frequencies.
Finally, SFL g-modes in rotating NSs could be excited 
due to gravitational driven (CFS) instability,
though this scenario 
does not seem very realistic \cite{lai99} 
because ${\mathcal N}$ is low,
which results in a large gravitational radiation timescale.

To conclude, we have predicted
a new class of g-modes in SFL NSs. 
We have calculated 
their Brunt-V$\ddot{\rm a}$is$\ddot{\rm a}$l$\ddot{\rm a}$ 
frequency $\mathcal{N}$, 
which strongly depends on $T$
and vanishes at $T=0$. 
The SFL g-modes appear to be unstable for certain temperature gradients 
(that correspond to $\mathcal{N}^2<0$). 
We have derived the criterion 
for convective instability
(analogue of the Schwarzschild criterion) 
in SFL NS cores. 
We have shown that
convection in the 
NS core may occur 
for both positive and negative temperature gradients
and is extremely sensitive to the model EOS and nucleon SFL.
We have only outlined the properties of SFL g-modes.
In particular, we have not calculated their frequency spectrum 
and damping times.
We plan to fill these gaps in the future publication.
Though we have only considered $npe$-matter of NSs,
our analysis can be easily extended to SFL hyperon and quark stars,
for which we also predict the existence 
of global, low-frequency SFL g-modes.

{\it Acknowledgements}. --- 
%
We are grateful to A.~I.~Chugunov and D.~G.~Yakovlev
for discussions and critical comments.
This study was supported 
by Ministry of Education and Science of Russian Federation 
(contract No.\ 11.G34.31.0001 
with SPbSPU and leading scientist G.G.\ Pavlov, 
and Agreement No.\ 8409, 2012), 
RFBR (grants 11-02-00253-a and 12-02-31270-mol-a), 
and by RF president programme 
(grants MK-857.2012.2 and NSh-4035.2012.2).

\bibliography{literature}

\end{document}